\documentclass[conference]{IEEEtran}
\IEEEoverridecommandlockouts
\usepackage{amsmath,amsfonts}
\usepackage{algorithm}
\usepackage{algpseudocode}
\usepackage{array}
\usepackage{tikz}
\usetikzlibrary{positioning}
\usepackage[caption=false,font=footnotesize,labelfont=sf,textfont=sf]{subfig}
\usepackage[colorlinks=true, linkcolor=red, citecolor=red, urlcolor=blue]{hyperref}
\usepackage{textcomp}
\usepackage{stfloats}
\usepackage{diagbox}
\usepackage{url}
\usepackage{verbatim}
\usepackage{graphicx}
\usepackage{cite}
\usepackage{bm}
\usepackage{amsthm}
\usepackage{proof}
\usepackage{amssymb}
\usepackage{color}
\usepackage{dsfont} 

\allowdisplaybreaks
\def\BibTeX{{\rm B\kern-.05em{\sc i\kern-.025em b}\kern-.08em
		T\kern-.1667em\lower.7ex\hbox{E}\kern-.125emX}}
\begin{document}
	
	\title{Environment-Aware Near-Field Channel Estimation Leveraging CKM and ISAC\\[0.3em]
		\large\itshape (Invited Paper)}
	
	\author{
		\IEEEauthorblockN{Yuan Guo$^{\dagger \S}$, Yilong Chen$^{\dagger \S}$, Zixiang Ren$^{\ddagger}$, and Jie Xu$^{\dagger \S}$}
		\IEEEauthorblockA{$^{\dagger}$School of Science and Engineering, The Chinese University of Hong Kong (Shenzhen)\\
			$^{\S}$Shenzhen Future Network of Intelligence Institute\\
			$^{\ddagger}$Department of Electrical and Computer Engineering, National University of Singapore\\
			Email: guoyuan@cuhk.edu.cn, yilongchen@link.cuhk.edu.cn, zixiang\_ren@nus.edu.sg, xujie@cuhk.edu.cn}
	}
	
	\maketitle

	\begin{abstract}
		This paper proposes an environment-aware near-field channel estimation framework for integrated sensing and communication (ISAC) systems equipped with extremely large-scale antenna arrays (ELAAs). The proposed framework jointly exploits channel knowledge maps (CKMs) and ISAC to obtain \emph{a priori} information on static and dynamic environmental features for facilitating channel estimation. In particular, we propose a novel CKM representation, termed the virtual object map (VOM), which stores the locations of virtual environment objects (EOs) to characterize the dominant multipath components (MPCs) induced by static physical EOs. In addition, we design a sensing-assisted channel training protocol, in which the ISAC-enabled base station (BS) transmits downlink pilots while simultaneously collecting monostatic echoes for sensing dynamic targets in the environment, and the user equipment (UE) feeds back a quantized version of its received pilot observation. Based on the VOM prior and the sensed dynamic information, the BS jointly estimates the coefficients of the static and dynamic MPCs to recover the near-field channel. Numerical results demonstrate that the proposed joint VOM- and sensing-aided channel estimation scheme significantly outperforms conventional schemes without VOM-based priors and/or dynamic sensing in terms of both channel estimation accuracy and achievable rate.
	\end{abstract}

	\section{Introduction}	
	Extremely large-scale antenna arrays (ELAAs) have emerged as a key enabling technology for sixth-generation (6G) wireless networks. By operating in the near-field regime, ELAAs provide enhanced beamforming capability and higher spatial multiplexing gains, thereby significantly improving data rates and transmission reliability \cite{liu2023near,cui9693928,wang2023near}. To fully exploit these advantages, accurate channel state information (CSI) is essential. However, conventional channel estimation methods rely on real-time pilot-based training, which fail to exploit environmental information and thus may incur prohibitive overhead for high-dimensional channels.

	Recently, channel knowledge maps (CKMs) and integrated sensing and communication (ISAC) have emerged as promising approaches for enabling environment-aware communications in 6G to address the above challenges \cite{wang2023near,Zeng10430216,Zeng9373011,Wu10287775,ren2024sensing}. In particular, wireless channels can be naturally decomposed into static and dynamic components \cite{Zeng10430216}. The static components are determined by persistent environmental structures such as buildings and walls, whereas the dynamic components arise from moving objects and transient scatterers. CKM and ISAC can be leveraged to acquire prior information of  these two components, respectively, and accordingly enhance the efficiency of channel estimation.  Specifically, CKM establishes a mapping from transmitter/receiver locations to channel knowledge, providing site-specific prior information about quasi-static environmental features, which can effectively reduce the uncertainty and overhead of online channel acquisition \cite{Zeng9373011,Zeng10430216,Wu10287775}. On the other hand, ISAC enables wireless transceivers to actively sense the environment via monostatic echoes, thereby capturing real-time variations of dynamic propagation components \cite{ren2024sensing}.

	In this paper, we investigate environment-aware near-field channel estimation for ELAA systems by jointly leveraging CKM and ISAC. We consider a downlink scenario where an ISAC-enabled BS  equipped with an ELAA serves  a single-antenna user equipment (UE). During transmission, the BS sends downlink pilots while simultaneously collecting monostatic echoes for sensing dynamic targets in the environment, and the UE feeds back a quantized version of its received pilot observations. To facilitate near-field channel estimation, we propose a novel CKM representation termed the virtual object map (VOM), which records the locations of virtual first-hop environmental objects (EOs) associated with static multi-path components (MPCs). The VOM enables the construction of near-field array responses associated with both sensing and communication channels. Leveraging the VOM, the BS first suppresses static clutter in the sensed echoes and extracts information about dynamic targets, and then obtains the near-field array responses of both static and dynamic MPCs for downlink communication. The BS subsequently estimates complex coefficients of these MPCs by using the quantized pilot feedback, thereby efficiently recovering the high-dimensional near-field channel vector. Numerical results demonstrate that the proposed scheme significantly outperforms conventional methods that do not exploit VOM priors or dynamic sensing, in terms of both channel estimation accuracy and achievable rate.

	\section{System Model}
	We consider a near-field ISAC system in which an ELAA-equipped BS serves a single-antenna UE. The BS is equipped with $N$ transmit/receive antenna elements. Let $\mathbf q_n\in\mathbb R^2$, $n\in\{1,\ldots,N\}$, denote the location of the $n$-th transmit/receive antenna pair, and let $\mathbf u\in\mathcal A\subset\mathbb R^2$ denote the UE location, with $\mathcal A$ denoting the two-dimensional region of interest (ROI). We use $\mathbf q_0=\frac{1}{N}\sum_{n=1}^{N}\mathbf q_n$ to denote the array reference point, and define the array aperture as $D=\max_{m,n\in\{1,\ldots,N\}}\|\mathbf q_m-\mathbf q_n\|$. We assume that the ROI $\mathcal A$ lies entirely in the radiative near-field region of the BS array, i.e., $\sqrt[3]{{D^4}/{8\lambda}} < \|\mathbf q_n-\mathbf u\| < {2D^2}/{\lambda}$, $\forall$$n\in\{1,\ldots,N\}$,  $\forall$$\mathbf u\in\mathcal A$, where $\lambda$ is the carrier wavelength \cite{liu2023near}.
		\begin{figure}[t]
		\centering
		\includegraphics[width=0.6\linewidth]{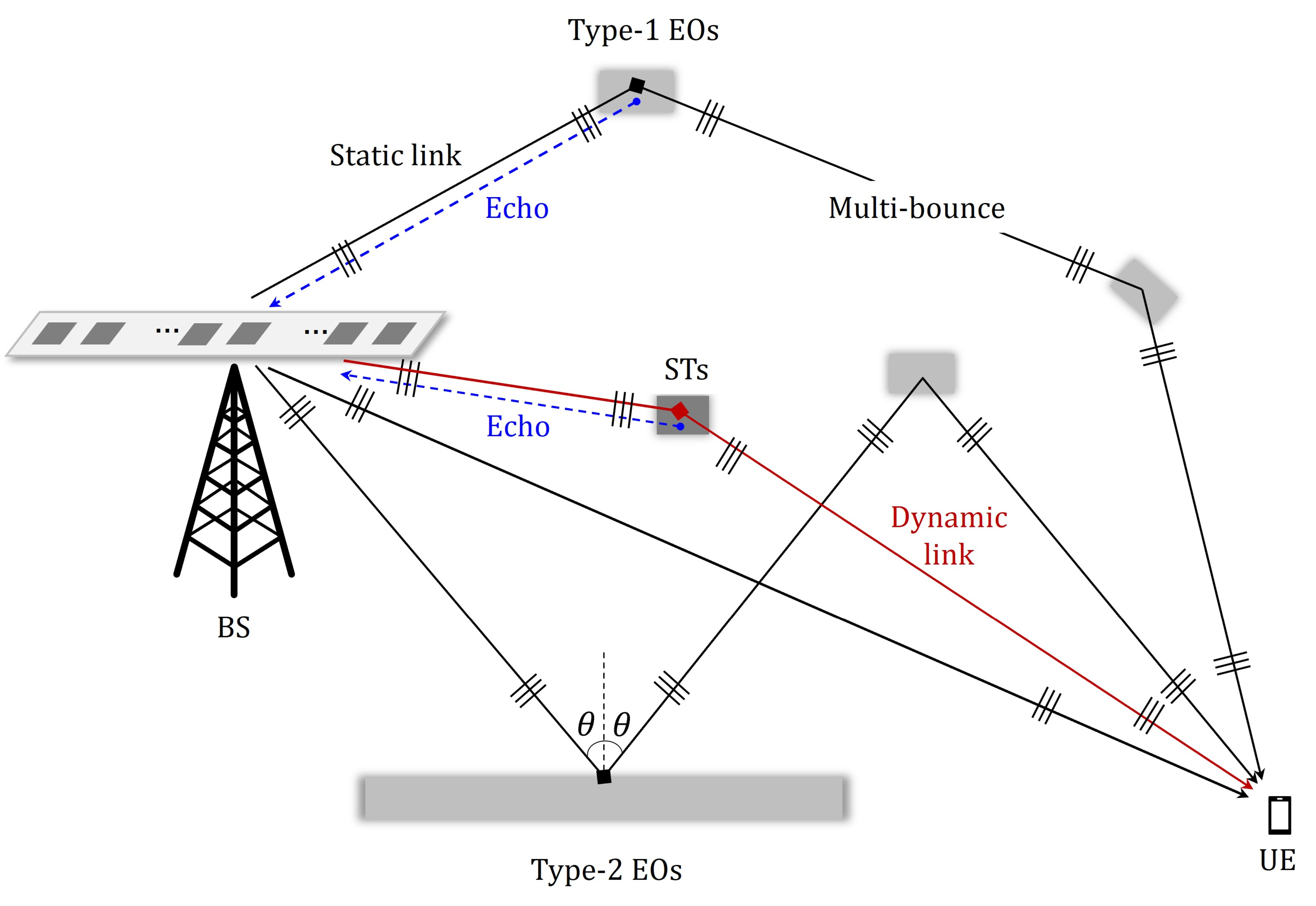}
		\caption{Illustration of the near-field ISAC channel  consisting of Type-1 and Type-2 EOs, as well as dynamic STs.}
		\label{fig:11}
		\vspace{-5mm}
	\end{figure}
	
	We consider the downlink channel estimation in a frequency-division duplexing (FDD) setting over quasi-static wireless environments, in which the downlink channel is acquired through downlink training and UE feedback \cite{ren2024sensing}, and the channel propagation environment is assumed to remain quasi-static within each coherence block. In particular, the downlink channel is modeled as a superposition of MPCs. Each MPC is characterized by a BS-visible  primary interaction location $\mathbf s$ and an associated complex coefficient. The location $\mathbf s$ denotes the first interaction point between the BS-emitted signal and an EO or ST, after which the propagated field travels toward the UE along the remaining path. Such  primary interaction locations may arise from three types of objects, namely,  Type-1 EOs, Type-2 EOs, and sensing targets (STs) \cite{3gpp38901v18}, as illustrated in Fig.~\ref{fig:11}. More specifically, Type-1 EOs are static compact objects with known locations whose interaction with the incident wave may involve reflection, scattering, or diffraction depending on the local geometry and material properties; Type-2 EOs are static large smooth surfaces with known locations and finite spatial extent, such as building facades, walls, or the ground, which mainly give rise to specular reflections; and STs represent dynamic target objects whose presence or geometry may vary across coherence blocks while remaining approximately static within each coherence block \cite{3gpp38901v18}. Accordingly, the downlink channel from the BS to a UE located at $\mathbf u$ is given by
	\begin{align}\label{eq:h_vector_model}
		\bm h(\mathbf u)=&
		\sum\limits_{\mathbf s\in\mathcal C^{\rm T1}\cup \mathcal C^{\rm T2}}
		\beta(\mathbf s,\mathbf u)\bm{\vartheta}(\mathbf s)
		+
		\sum\limits_{\mathbf s\in\mathcal C^{\rm ST}}
		\beta(\mathbf s,\mathbf u)\bm{\vartheta}(\mathbf s),
	\end{align}
	where $\mathcal C^{\rm T1}$, $\mathcal C^{\rm T2}$, and $\mathcal C^{\rm ST}$ denote the sets of BS-visible primary interaction points induced by Type-1 EOs, Type-2 EOs, and STs, respectively; $\beta(\mathbf s,\mathbf u)\in\mathbb C$ denotes the corresponding complex path coefficient, that may change dynamically over blocks, with $\beta(\mathbf s,\mathbf u)=0$ whenever $\mathbf s$ does not contribute to the channel toward $\mathbf u$. Furthermore,  $\bm{\vartheta}(\mathbf p)\in\mathbb C^{N\times 1}$ denotes the near-field array response from the BS to point $\mathbf p\in\mathbb R^2$, given by
	\vspace{-3mm}
	\begin{equation}\label{eq:nf_array_response_compact}
		\bm{\vartheta}(\mathbf p)
		=
		\frac{\lambda}{4\pi}
		\left[
		\frac{e^{-j k \|\mathbf q_1-\mathbf p\|}}{\|\mathbf q_1-\mathbf p\|},
		\ldots,
		\frac{e^{-j k \|\mathbf q_N-\mathbf p\|}}{\|\mathbf q_N-\mathbf p\|}
		\right]^T,
	\end{equation}
	with $k=\frac{2\pi}{\lambda}$. 
	
	Similarly, the monostatic sensing channel from the BS transmit array to its receive array can be modeled using the same primary interaction points, where the outgoing and returning paths share the same interaction point. Accordingly, the sensing channel, denoted by $\bm H\in\mathbb{C}^{N\times N}$, is given by
	\begin{align}\label{eq:round_trip_channel}
		\hspace{-3mm}	\bm H
		=
		\sum_{\mathbf s\in\mathcal C^{\rm T1}\cup \mathcal C^{\rm T2}}
		\tilde{\gamma}(\mathbf s)\bm\vartheta(\mathbf s)\bm\vartheta^{T}(\mathbf s)
		+
		\sum_{\mathbf s\in\mathcal C^{\rm ST}}
		\tilde{\gamma}(\mathbf s)\bm\vartheta(\mathbf s)\bm\vartheta^{T}(\mathbf s),
	\end{align}
	where $\tilde{\gamma}(\mathbf s)\in\mathbb C$ denotes the corresponding complex round-trip coefficient.  
	
	Finally, in the considered system, both the downlink communication and monostatic sensing channels contain static and dynamic components. In the downlink channel, the MPCs induced by Type-1 and Type-2 EOs form the static part, while those induced by STs form the dynamic part. In the sensing channel, the background echoes from Type-1 and Type-2 EOs are persistent across coherence blocks, whereas those from STs capture block-dependent variations. Since the static components are environment-dependent and stable over many coherence blocks, they can be represented as reusable priors in the VOM, as detailed in the next section.

	\section{Virtual Object Map (VOM)}
	
	To capture reusable environment-dependent priors for channel estimation, we introduce a VOM. Rather than storing the static physical EOs themselves, the VOM is built upon a finite set of virtual objects induced by these EOs, namely a common virtual-object library
	\begin{equation}\label{eq:vom_location_library}
		\mathcal V=\{\mathbf s_1,\ldots,\mathbf s_L\},
	\end{equation}
	where each virtual object is represented by a BS-visible primary interaction location $\mathbf s_\ell$ induced by a static Type-1 or Type-2 EO.
	
	Based on the common virtual-object library $\mathcal V$, the VOM establishes a location-to-index mapping. More specifically, for a UE location $\mathbf u\in\mathcal A$, the VOM maps $\mathbf u$ to an index set
	\begin{equation}\label{eq:vom_map_u}
		\mathcal F(\mathbf u)\subseteq\{1,\ldots,L\},
	\end{equation}
	which contains the indices of the $J$ dominant virtual objects in $\mathcal V$, i.e. $|\mathcal F(\mathbf u)|=J$, contributing to the long-term static BS-to-UE channel. Specifically, the virtual objects in $\mathcal V$ are ranked according to the long-term contribution metric given by
	\begin{equation}\label{eq:vom_metric_u}
		m(\mathbf s_\ell,\mathbf u)=\mathbb E\!\left[|\beta(\mathbf s_\ell,\mathbf u)|^2\right]\|\bm\vartheta(\mathbf s_\ell)\|_2^2,
	\end{equation}
	and the indices of the top-$J$ virtual objects are stored in $\mathcal F(\mathbf u)$.
	
	Furthermore, note that monostatic sensing can be viewed as a special case when the location argument $\mathbf u$ coincides with the BS location $\mathbf q_0$. In this case, the VOM maps $\mathbf q_0$ to an index set
	\begin{equation}\label{eq:vom_map_q0}
		\mathcal F(\mathbf q_0)\subseteq\{1,\ldots,L\},
	\end{equation}
	which contains the indices of the $K$ dominant virtual objects in $\mathcal V$, i.e., $|\mathcal F(\mathbf q_0)|=K$, contributing to the long-term static monostatic background. Specifically, the virtual objects in $\mathcal V$ are also ranked according to
	\begin{equation}\label{eq:vom_metric_q0}
		\rho(\mathbf s_\ell)=\mathbb E\left[|\tilde{\gamma}(\mathbf s_\ell)|^2\right]\|\bm\vartheta(\mathbf s_\ell)\|_2^4,
	\end{equation}
	and the indices of the top-$K$ virtual objects are stored. 
	
	In practice, the VOM is constructed offline based on site-specific environment information. In particular, a ray-tracing solver is used to identify BS-visible virtual object locations induced by static Type-1 and Type-2 EOs, and spatial clustering is then applied to form the common virtual-object library  $\mathcal V$ \cite{Wu10720902}. For each sampled UE location, the corresponding index set $\mathcal F(\mathbf u)$ is obtained by ranking the virtual-object locations in $\mathcal V$ according to \eqref{eq:vom_metric_u} and retaining the top-$J$ ones. Similarly, the BS-side index set $\mathcal F(\mathbf q_0)$ is obtained according to \eqref{eq:vom_metric_q0}.

	\begin{figure}[t] \centering \includegraphics[width=1\linewidth]{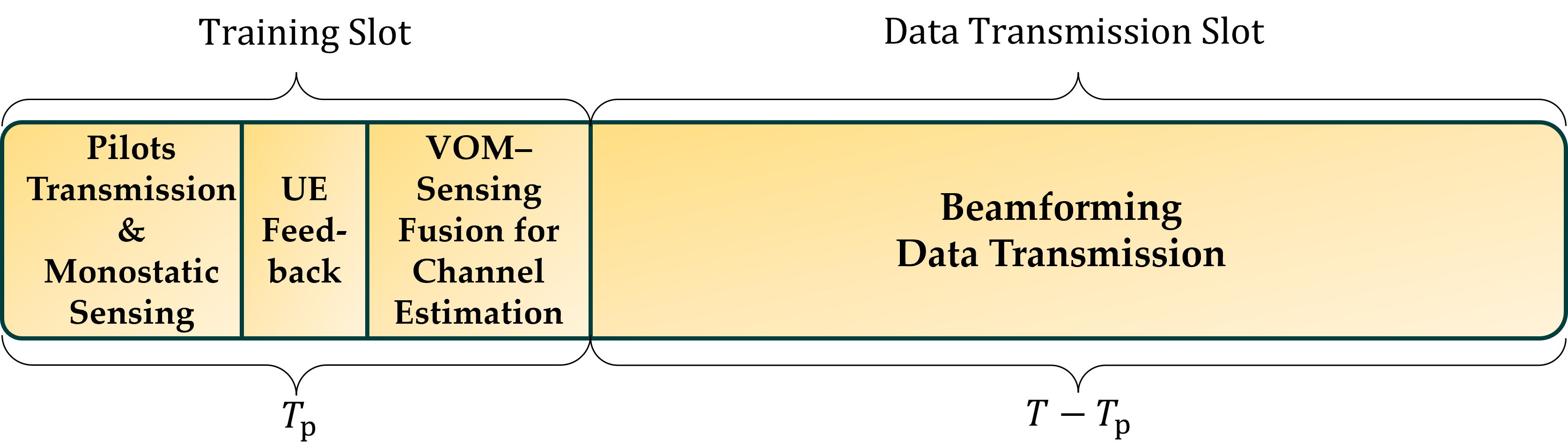}\vspace{-3mm} \caption{Overall protocol of the proposed joint VOM- and sensing-aided near-field channel estimation and data transmission framework.} \label{fig:protocolfig}
		\vspace{-5mm} \end{figure}

	\section{Joint VOM- and Sensing-assisted Near-field Channel Estimation}\label{sec:vom_recon}
	In this section, we present the proposed environment-aware sensing-assisted near-field channel estimation framework by leveraging VOM. We first introduce the operation protocol and then develop the corresponding channel estimation algorithm.
	\subsection{Operation Protocol}
	
	We consider that each coherence block of $T$ symbols is divided into two slots, with $T_{\rm p}$ symbols for training and $T-T_{\rm p}$ symbols for data transmission, respectively, as illustrated in Fig.~\ref{fig:protocolfig}.

	During the first slot, the BS transmits a known pilot matrix $\bm Z\in\mathbb C^{N\times T_{\rm p}}$, and the resulting pilot observation at the UE is given by
	\begin{equation}\label{eq:pilot_obs_protocol}
		\bm y
		=
		\bm Z^{H}\bm h(\mathbf u)
		+
		\bm n,
	\end{equation}
	where $\bm n\in\mathbb C^{T_{\rm p}\times 1}$ is the additive white Gaussian noise (AWGN) vector following a circularly symmetric complex Gaussian distribution (CSCG) with mean $\bm 0$ and covariance matrix $\sigma^2\bm I_{T_{\rm p}}$, i.e., $\bm n\sim\mathcal{CN}(\bm 0,\sigma^2\bm I_{T_{\rm p}})$. The UE subsequently feeds back a quantized version of its received pilot observation to the BS. To reduce the uplink feedback overhead, we adopt a  quantized feedback architecture based on a shared codebook. Specifically, let $\mathcal C=\{\bm c_1,\ldots,\bm c_{2^B}\}$ denote the pre-designed quantization codebook shared by the BS and the UE. The UE selects the codeword index $\hat b$ that best represents its pilot observation and feeds it back using $B$ bits, in line with the standard limited-feedback architecture \cite{Love2008}. Upon receiving $\hat b$, the BS reconstructs the quantized pilot observation as $\hat{\bm y}=\bm c_{\hat b}$ for subsequent channel estimation. To focus on the performance of downlink channel estimation, we assume an ideal error-free feedback link so that the quantized pilot observation is perfectly recovered at the BS \cite{ren2024sensing,sohrabi2021deep}. Moreover, the UE feedback and the subsequent BS-side channel estimation are assumed to incur negligible latency compared with the coherence block duration, and thus are not explicitly counted in the symbol-level overhead.

	Furthermore, during the training slot, the BS simultaneously collects the monostatic echo signals, which are given by
	\begin{equation}\label{eq:echo_obs_protocol}
		\bm E
		=
		\bm H\bm Z
		+
		\bm N,
	\end{equation}
	where $\bm N\in\mathbb C^{N\times T_{\rm p}}$ denotes the sensing noise matrix with each entry following independent and identically distributed (i.i.d.)  CSCG distribution $\mathcal{CN}(0,\sigma_{\rm s}^{2})$. 	The echo observation $\bm E$ is exploited to extract information about the ST-induced propagation components, thereby providing dynamic environmental information to assist downlink channel estimation. By jointly leveraging the VOM prior, the monostatic echo observation, and the reconstructed quantized pilot observation, the BS estimates the downlink channel, as detailed in the following subsections. The resulting estimated channel  is then used to design the downlink beamforming vector for data transmission in the second slot, e.g., via maximum-ration transmission (MRT) to maximize the downlink signal-to-noise ratio (SNR).
	
	In the following, we first introduce the extraction of dynamic ST-related propagation information from the monostatic echo observation, and then present the downlink channel estimation procedure based on the quantized pilot observation by jointly leveraging the VOM prior and the sensing information.
	
	\subsection{VOM-Aided Dynamic Information Extraction}\label{subsec:dynamic_extract}
	
	We now describe how to extract dynamic ST-related propagation information from the monostatic echo observation $\bm E$ in \eqref{eq:echo_obs_protocol}. To this end, the VOM is used to characterize the dominant static virtual objects and suppress their contributions to $\bm E$, such that the residual observation is mainly determined by the dynamic ST-induced signals. Recall that
	$
	\mathcal F(\mathbf q_0)=\{\ell_1,\ldots,\ell_K\}
	$
	denotes the BS-side VOM mapping for the monostatic sensing channel, $\{\hat{\mathbf s}_{\ell_1}, \ldots, \hat{\mathbf s}_{\ell_K}\}$ denote the corresponding locations of the $K$ virtual objects. We define   the corresponding near-field response matrix as
	\[
\bm A_{\mathrm{sens}}
=
\big[
\bm\vartheta(\hat{\mathbf s}_{\ell_1}),\ldots,\bm\vartheta(\hat{\mathbf s}_{\ell_K})
\big].
	\]
	
	Based on $\bm A_{\mathrm{sens}}$, we construct the subspace associated with the dominant static echoes, which is then used for static clutter suppression. Specifically, let $\bm A_{\mathrm{sens}}=\bm U_{\mathrm{sens}}\bm R_{\mathrm{sens}}$ be the thin QR decomposition of $\bm A_{\mathrm{sens}}$, where the columns of $\bm U_{\mathrm{sens}}$ form an orthonormal basis for this subspace. The clutter-suppressed echo is then given by
	\begin{equation}\label{eq:proj_echo}
		\tilde{\bm E}
		=
		\bm P^{\perp}\bm E,
	\end{equation}
	where $\bm P^{\perp}=\bm I-\bm U_{\mathrm{sens}}\bm U_{\mathrm{sens}}^H$ is the orthogonal projector onto the complement of the subspace spanned by $\bm U_{\mathrm{sens}}$.

	After static clutter suppression, the projected echo $\tilde{\bm E}$ is mainly induced by the dynamic STs. To accommodate a general dynamic-target model, we consider the case where the ST-induced echo may be spatially extended, such that its energy is distributed over a localized region rather than concentrated on a few isolated points. In this case, instead of performing explicit point localization, we extract a low-dimensional array-domain subspace from $\tilde{\bm E}$. Applying the economy singular value decomposition gives
	$
	\tilde{\bm E}
	=
	\bm U_{\mathrm e}\bm\Sigma_{\mathrm e}\bm V_{\mathrm e}^{H}.
	$
	Let $\sigma_1\ge\sigma_2\ge\cdots$ denote the singular values. We choose the subspace dimension as
	\begin{equation}\label{eq:ext_r_select}
		\varrho
		=
		\min
		\bigg\{
		\varrho_{\max},
		\min\bigg\{
		k:
		\frac{\sum_{i=1}^{k}\sigma_i^2}{\sum_i \sigma_i^2}\ge \eta
		\bigg\}
		\bigg\},
	\end{equation}
	where the ratio measures the cumulative energy captured by the first $k$ dominant singular components of $\tilde{\bm E}$. Accordingly, $\varrho$ is chosen as the smallest integer such that the retained subspace captures at least an $\eta$ fraction of the total echo energy, subject to the upper bound $\varrho_{\max}$. The parameter $\varrho_{\max}$ is introduced to limit the dimension of the extracted dynamic subspace, thereby controlling the subsequent channel estimation complexity and avoiding the inclusion of weak noise-dominated components. We then define the resulting dynamic basis as
	$
	\tilde{\bm U}_{\mathrm e}
	=
	\bm U_{\mathrm e}(:,1:\varrho).
	$
	
	\subsection{Downlink Channel Estimation}\label{subsec:joint_recon}
	
	Finally, with the VOM-specified static virtual objects and the sensing-derived dynamic subspace obtained, we estimate the overall downlink channel $\bm h(\mathbf u)$. Similarly, let
	$
	\mathcal F(\mathbf u)=\{\ell_1,\ldots,\ell_J\}
	$
	denote the VOM mapping for the downlink communication channel at user location $\mathbf u$, and let
	$
	\bar{\mathcal C}(\mathbf u)
	=
	\{\bar{\mathbf s}_{\ell_1},\ldots,\bar{\mathbf s}_{\ell_J}\}
	$
	denote the corresponding locations of the $J$ static virtual objects. We then define the associated near-field response matrix as
	$
	\bm A_{\mathrm{sta}}(\mathbf u)
	=
	\big[
	\bm\vartheta(\bar{\mathbf s}_{\ell_1}),\ldots,\bm\vartheta(\bar{\mathbf s}_{\ell_J})
	\big].
	$ 
	Using the VOM-retrieved static virtual objects and the dynamic subspace extracted from the monostatic echo signal, the downlink channel is approximated as
	\begin{align}\label{eq:h_model_joint}
		\bm h(\mathbf u)
		\approx&
		\sum\nolimits_{\mathbf s\in\bar{\mathcal C}(\mathbf u)}
		\beta(\mathbf s,\mathbf u)\bm{\vartheta}(\mathbf s)
		+
		\tilde{\bm U}_{\mathrm e}\bm\xi \nonumber\\
		=&
		\bm A_{\mathrm{sta}}(\mathbf u)\bm\alpha(\mathbf u)
		+
		\tilde{\bm U}_{\mathrm e}\bm\xi,
	\end{align}
	where $\bm\alpha(\mathbf u)\in\mathbb C^{J\times 1}$ collects the coefficients associated with the VOM-retrieved static virtual objects, and $\bm\xi\in\mathbb C^{\varrho\times 1}$ denotes the dynamic subspace coefficient vector. In addition, the quantized pilot observation is further expressed as
$
		\hat{\bm y}
		\approx
		\bm Z^H \bm A_{\mathrm{sta}}(\mathbf u)\bm\alpha(\mathbf u)
		+
		\bm Z^H \tilde{\bm U}_{\mathrm e}\bm\xi
		+
		\bm n.
$
	The unknown vectors $\bm\alpha(\mathbf u)$ and $\bm\xi$ are jointly estimated via regularized least squares, i.e.,
	\begin{align}\label{eq:joint_ridge}
		(\bm\alpha^{\star},\bm\xi^{\star})
		=&
		\arg\min_{\bm\alpha,\bm\xi}\Big(
		\big\|
		\hat{\bm y}
		-
		\bm Z^H \bm A_{\mathrm{sta}}(\mathbf u)\bm\alpha
		-
		\bm Z^H \tilde{\bm U}_{\mathrm e}\bm\xi
		\big\|_2^2 \nonumber\\
		&\hspace{28mm}+
		\mu_{\mathrm s}\|\bm\alpha\|_2^2
		+
		\mu_{\mathrm d}\|\bm\xi\|_2^2\Big),
	\end{align}
	where $\mu_{\mathrm s}>0$ and $\mu_{\mathrm d}>0$ denote the regularization parameters for $\bm\alpha$ and $\bm\xi$, respectively, which are introduced to improve robustness against noise and possible ill-conditioning under limited pilot observations. Let $\bm\Phi_{\mathrm s}=\bm Z^H \bm A_{\mathrm{sta}}(\mathbf u)$ and $\bm\Phi_{\mathrm d}=\bm Z^H \tilde{\bm U}_{\mathrm e}$, and define $\bm\Phi=[\bm\Phi_{\mathrm s},\bm\Phi_{\mathrm d}] $.
	Then, the solution to \eqref{eq:joint_ridge} is given by
	\begin{equation}\label{eq:joint_ridge_sol}
		\bm z^{\star}
		=
		\left(
		\bm\Phi^H\bm\Phi
		+
		\begin{bmatrix}
			\mu_{\mathrm s}\bm I & \bm 0\\
			\bm 0 & \mu_{\mathrm d}\bm I
		\end{bmatrix}
		\right)^{-1}
		\bm\Phi^H\hat{\bm y},
	\end{equation}
	where $
	\bm z^{\star}
	=
	\begin{bmatrix}
		\bm\alpha^\star\\
		\bm\xi^\star
	\end{bmatrix}.
	$ The resulting channel estimate is thus given by
	\begin{equation}\label{eq:h_final_hat}
		\hat{\bm h}(\mathbf u)
		=
		\bm A_{\mathrm{sta}}(\mathbf u)\bm\alpha^{\star}
		+
		\tilde{\bm U}_{\mathrm e}\bm\xi^{\star}.
	\end{equation}

	It follows from \eqref{eq:h_final_hat} that the channel is reconstructed as the sum of a VOM-specified static component and a sensing-derived dynamic component, rather than estimated as an unstructured whole from the pilot observation alone. This structured formulation reduces the effective search space and enables the limited pilot observation to focus on estimating the coefficients over two physically meaningful bases. The same formulation also accommodates point-like dynamic STs, for which the sensing-derived component becomes a rank-one or, more generally, a low-dimensional special case.
	
	\section{Simulation Results}
	\begin{figure}
		\centering
		\includegraphics[width=0.68\linewidth]{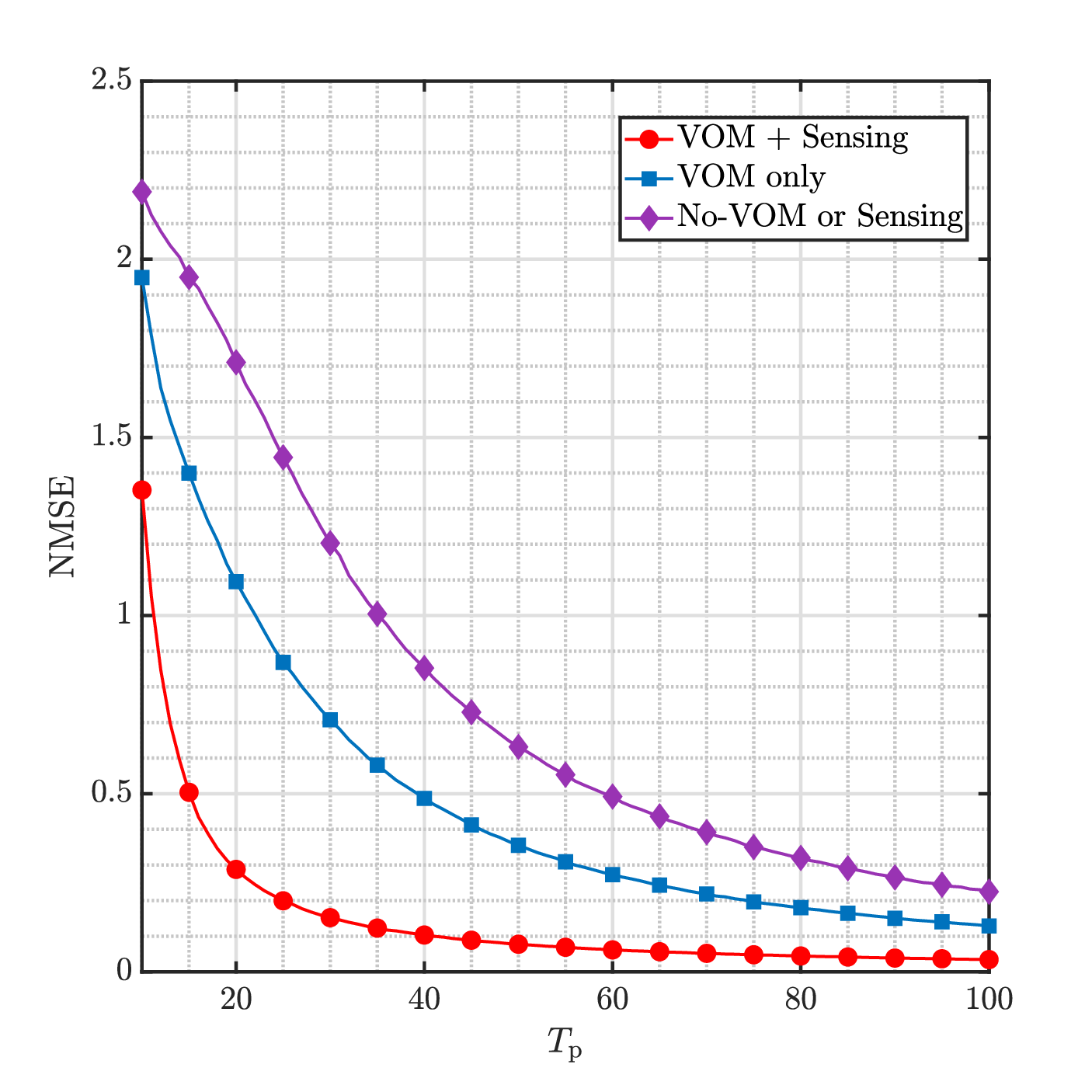}
		\vspace{-5mm}
		\caption{NMSE versus pilot length $T_{\rm p}$.}
		\label{fig:nmse}
		\vspace{-5mm}
	\end{figure}
	In this section, we evaluate the proposed joint VOM- and sensing-aided near-field channel estimation framework. Unless otherwise specified, we consider a $64$-antenna ULA at $2.4$~GHz with half-wavelength spacing, yielding a Rayleigh distance of $248.1$~m, a UE located at $(0,20)$~m, and a coherence block length of $T=400$. The VOM communication and sensing entries contain $J=5$ and $K=20$ dominant virtual objects, respectively. The downlink pilot, echo, and data transmission SNRs are set to $5$~dB, $40$~dB, and $10$~dB, respectively. Unless otherwise specified, the static EOs are uniformly distributed in the rectangular region $[-7,7]\times[5,15]$~m. The dynamic ST is modeled as a circular cluster centered at $(-1.5,5)$~m with radius $1.5$~m, and the dynamic subspace parameters are set as $\varrho_{\max}=5$ and $\eta=0.9$. We compare the proposed scheme with the following benchmarks:
	\begin{itemize}
		\item \textbf{Benchmark design with (w/) VOM, but without (w/o) sensing}: The VOM provides the static channel basis, while the dynamic component is represented by a few atoms selected from the polar-domain codebook after projection onto the orthogonal complement of the corresponding static measurement subspace; the channel is then estimated by joint coefficient estimation.
		\item \textbf{Conventional design w/o VOM or sensing}: The whole channel is estimated directly from the pilot observation using the near-field polar-domain codebook in \cite{cui9693928}.
	\end{itemize}
	The performance is evaluated in terms of the normalized mean squared error (NMSE) of channel estimation and the achievable downlink rate under MRT based on the estimated channel. Specifically, the NMSE is defined as
	$
	\mathrm{NMSE}
	=
	{\|\bm h(\mathbf u)-\hat{\bm h}(\mathbf u)\|_2^2}/
	{\|\bm h(\mathbf u)\|_2^2},
	$
	and the achievable downlink rate is given by
	$
	R
	=
	\left(1-{T_{\rm p}}/{T}\right)
	\log_2
	\left(
	1+
	{P\,|\bm h^H(\mathbf u)\bm w|^2}/{\sigma^2}
	\right),
	$
	where $\bm w=\hat{\bm h}(\mathbf u)/\|\hat{\bm h}(\mathbf u)\|_2$.

	\begin{figure}
		\centering
		\includegraphics[width=0.68\linewidth]{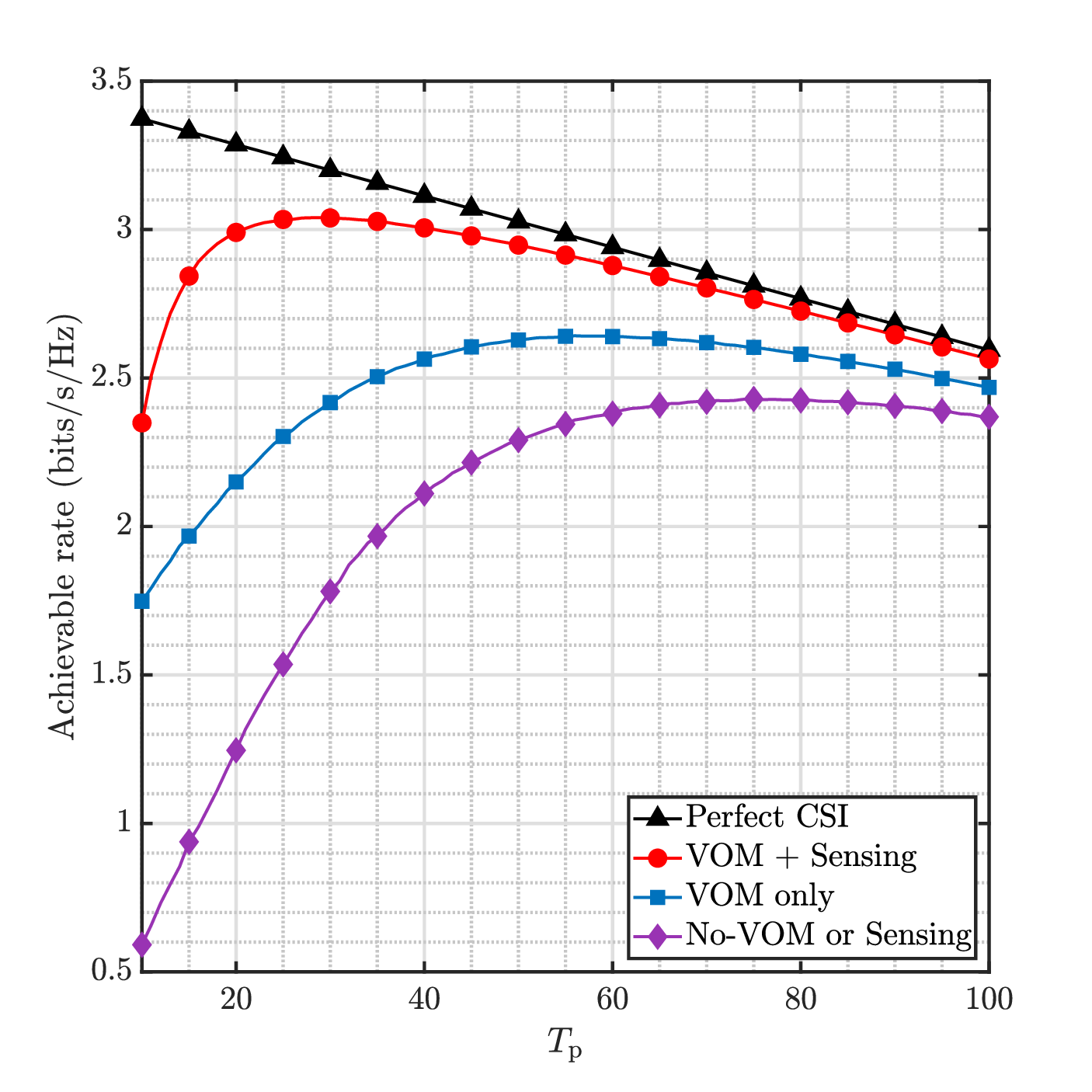}
		\vspace{-5mm}
		\caption{Achievable downlink rate versus pilot length $T_{\rm p}$.}
		\label{fig:rate}
		\vspace{-5mm}
	\end{figure}
	Fig.~\ref{fig:nmse} shows the NMSE of channel estimation versus the pilot length $T_{\rm p}$ for different schemes. It is observed that the proposed joint VOM- and sensing-aided scheme consistently achieves the lowest NMSE over the whole range of $T_{\rm p}$, followed by the VOM-only benchmark, while the scheme without VOM or sensing performs the worst. This confirms the effectiveness of exploiting environment-dependent priors for channel estimation. In particular, the VOM-only benchmark provides a clear gain over the scheme without VOM or sensing, because the VOM constrains the dominant static virtual objects and thus significantly reduces the uncertainty of the static channel component. Building on this, the proposed scheme further exploits monostatic sensing observations to extract the dynamic echo subspace after static clutter suppression, thereby enabling a more accurate recovery of the dynamic channel component. It is also observed that the NMSE gap is most pronounced in the short-pilot regime, which indicates that the proposed scheme is especially effective when pilot resources are limited.

	Fig.~\ref{fig:rate} shows the achievable downlink rate versus the pilot length $T_{\rm p}$, where the perfect-CSI curve corresponds to the ideal upper bound with exact downlink channel knowledge at the BS. Consistent with the normalized NMSE results in Fig.~\ref{fig:nmse}, the proposed joint VOM- and sensing-aided scheme achieves the highest rate among all schemes and remains close to the perfect-CSI upper bound, especially for moderate and large $T_{\rm p}$. This demonstrates that the improved channel estimation accuracy brought by jointly exploiting VOM priors and sensing information translates directly into beamforming gain. The VOM-only benchmark also consistently outperforms the scheme without VOM or sensing, which shows that static environment knowledge alone already provides substantial performance improvement even without dynamic sensing support. Moreover, the achievable rates by all schemes first increase and then decrease with $T_{\rm p}$. The reason is that increasing $T_{\rm p}$ improves channel estimation quality in the short-pilot regime, whereas, for large $T_{\rm p}$, the benefit of improved estimation is eventually outweighed by the increased pilot overhead. 
	\section{Conclusion}
	In this paper, we proposed a novel joint VOM- and sensing-aided framework for near-field downlink channel estimation in ELAA-enabled ISAC systems. The proposed VOM provides reusable environment-dependent priors in the form of dominant static virtual objects, based on which the static channel component is represented over a compact location-dependent basis. Meanwhile, after suppressing the static clutter in the sensed echoes, the BS extracts a low-dimensional dynamic subspace to capture the ST-induced channel component. By jointly exploiting the VOM prior, the sensing-derived dynamic information, and the quantized pilot feedback, the proposed framework enables accurate and low-overhead near-field channel estimation. Numerical results showed that the proposed framework achieves significant gains over benchmark schemes without VOM priors and/or sensing information in terms of both NMSE and achievable downlink rate. Future work may investigate adaptive VOM construction and online map updating in evolving environments.

	\bibliographystyle{IEEEtran}
	\bibliography{ref}
	
\end{document}